# Data integration in systems genetics and aging research

Alexis Rapin
Laboratory of Integrative Systems Physiology
EPFL
Lausanne, Switzerland
alexis.rapin@epfl.ch,
https://orcid.org/0000-0003-3448-5459

Maroun Bou Sleiman
Laboratory of Integrative Systems Physiology
EPFL
Lausanne, Switzerland
maroun.bousleiman@epfl.ch
https://orcid.org/0000-0002-1375-7577

Johan Auwerx
Laboratory of Integrative Systems Physiology
EPFL
Lausanne, Switzerland
johan.auwerx@epfl.ch
https://orcid.org/0000-0002-5065-5393

*Abstract*—Human life expectancy has dramatically improved over the course of the last century. Although this reflects a global improvement in sanitation and medical care, this also implies that more people suffer from diseases that typically manifest later in life, like Alzheimer and atherosclerosis. Increasing **healthspan** by delaying or reverting the development of these age-related diseases has therefore become an urgent challenge in biomedical research. Research in this field is complicated by the multi-factorial nature of age-related diseases. They are rooted in complex physiological mechanisms impacted by heritable, environment and life-style factors that can be unique to each individual. Although technological advances in high-throughput biomolecular assays have enabled researchers to investigate individual physiology at the molecular level, integrating information about its different components, and accounting for individual variations remains a challenge. We are using a large collection of "omics" and phenotype data derived from the **BXD mouse genetic diversity panel** to explore how good data management practices, as fostered by the **FAIR principles**, paired with an **explainable artificial intelligence** framework, can provide solutions to decipher the complex roots of age-related diseases. These developments will help to propose innovative approaches to extend healthspan in the aging global population.

## I. INTRODUCTION

Age-related diseases are associated to a number of heritable, environmental and life-style factors that impact physiology. In this context, the **systems genetics** community investigates the links between genetics, metabolism and individuals' traits to discover underlying molecular mechanisms, which could be utilized to design therapies and treatments. In the context of aging, the aim would be to reverse its effect and delay the development of its associated disorders. Systems genetics leverages the high-throughput capacity of "omics" technologies coupled with the sample availability and controlled experimental conditions offered by model organisms to assess biomolecular mechanisms in cells and tissues. While generating enormous amount of biomolecular data, the research community generally focuses on assessing the links between pairs of biological layers, such as associations between genetics and phenotypes, or between genetics and gene expression. Although this approach allows to reveal associations between individual factors, it hardly addresses interactions between more than two factors and patterns involving multiple tissues and multiple layers of physiological regulation. In contrast, there is increasing evidence that essential mechanisms underlying complex disorders can only be unveiled by considering multiple layers of biological observations together [1].

Despite this, only few attempts to integrate and analyze diverse biomedical data as a whole have been attempted. Indeed, both the construction of integrated **knowledge bases**, and the subsequent application of analysis methods, are technically challenging. Here, we summarize the main challenges for data integration in biomedicine, highlight trends and describe our current effort to overcome these challenges.

## II. AGE-RELATED DISEASES

### A. The leading causes of premature death

As the global population grows, and life expectancy increases, so does the number of people at risk for developing age-related diseases. Indeed, advances in sanitation, medicine and food security have contributed to considerably reduce child mortality and expand lifespan globally along the course of the last century. Subsequently, improving the health of the elderly and extending the so-called healthspan has emerged as a new challenge in medicine, as the leading causes of premature death moved from infections to cardiovascular diseases and cancer [2].

### B. Multiple risk factors

Age-related diseases represent a large spectrum of disorders, including neurodegenerative, cardiovascular and musculoskeletal diseases as well as cancer. The development of these disorders is typically multi-factorial. Along with age, important risk factors include genetics, diet, life-style, smoke and environmental exposures, as well as one's history of diseases and medication. In addition, intricate factors, such as the accumulation of epigenetic changes throughout life, referred to as the epigenetic "clock", the microbiome, and interactions between factors, are also important (Fig.1) [3]–[6]. As the unique combination of risk factors certainly differs from one patient to another, individual variations need to be taken into account both in research and the clinic. For example, studies in mouse have shown that the effect of preventive interventions aiming at extending lifespan, such as dietary restriction, can vary from beneficial to detrimental depending on one's genetic makeup [7]. Similarly, the composition of the intestinal microbiome has been shown to affect how food is absorbed and metabolized [8], [9].

At the cellular level, aging is characterized by the loss of intracellular proteostasis, mitochondrial homeostasis, and epigenetic alterations (Fig. 1) [10]. Identifying biomolecular pathways that can be exploited to slow down, delay or reverse these biological aging processes is therefore critical to address the leading health threats of today and tomorrow. This requires the investigation of physiological mechanisms at the molecular level across organs and tissues, while taking inter-individual variations into account.

## III. PRECISION MEDICINE

### A. Research approach

Precision – or personalized – medicine addresses complex diseases by adapting therapeutic approaches to the individual characteristics of the patient, and in particular to the genetics. This approach has been unlocked by the development of high-throughput biomolecular assays, or "omics", technologies, which allow to asses individuals physiology at the molecular level by drawing biomolecular profiles of tissues. From the research perspective, investigating the physiological mechanisms underlying complex conditions demands to profile numerous tissues, if not single cells, from a large diversity of subjects across multiple experimental conditions [11]. Such comprehensive research cannot be easily carried out in humans, mainly because the access to samples, and the control over important factors such as genetics, diet or environmental exposures, is limited. Indeed, although epidemiological studies can provide insights into the role of many factors that can reasonably be measured in human settings, such as genetics, clinical phenotypes and environmental factors, true experimentation to decipher of the underlying biomolecular mechanisms requires the use of experimental models. Relevant models range from cell lines and nematodes to larger organisms like mammals, depending on the question at hand.

### B. Mouse genetic diversity panels

In order to study complex systems, researcher's strategy is to control as many variables as can be while measuring as many of those that cannot be controlled and simultaneously inducing controlled variations of one or multiple variables. With this regards, mouse genetic diversity panels are a model of choice to assess the links between genetic variations and physiological traits associated with complex conditions, and are seen as the experimental counterpart of **precision medicine** [12]. Indeed, these panels are composed of genetically diverse inbred strains of mice and are designed to provide a stable and reproducible genetic diversity across cohorts. This model therefore allows to introduce defined genetic variations while simultaneously controlling environmental conditions and diet while providing access to a large variety of biological samples and enabling the measurement of a variety of phenotypes.

### C. Translational research

Although most of our knowledge in fundamental biology comes from model organisms, there are undeniable differences between human and mouse physiology. The research process in precision medicine is therefore a continuous cycle driven by epidemiological observations leading to the design of experiments on model organisms, and which results further demand validation in human settings [13]. Eventually, this translational process can lead to the prioritization and design of further human studies.

## IV. SYSTEMS GENETICS

### A. Associative studies

Linking genetic variations to phenotypes or to biomolecular profiles of tissues has been classically carried on in so called systems genetics studies using associative approaches, such as Genome-Wide Association Studies (GWAS) and Quantitative Trait Locus (QTL) mapping (Fig.

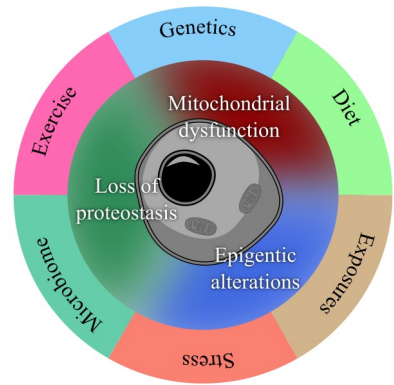

Fig. 1. Cellular aging is associated to mitochondrial dysfunction, the loss of proteostasis and epigenetic alteration and impacted by heritable, lifestyle and environmental factors.

2 A). These approaches screen "omics" data for any association with phenotypes, gene variations or any observation on other "omics" layers [14]. Although these approaches unveiled numerous insights into the genetic roots of complex diseases, they may not exploit the full potential of the multi-modal "omics" datasets that can be collected. Indeed, such associative approaches are limited to assessing the links between pairs of biological layers, such as between genetics and phenotypes, or between gene expression and phenotypes (Fig. 2 A). They are therefore bound to catch the "low-hanging fruits": single factors clearly associated with phenotypes (e.g. the genetic variant A is associated with a high body mass). This can miss important patterns of interactions across multiple layers of biological observations [15]: what if the genetic variant A was associated with a high body mass, but only if gene B is highly expressed in the liver while bacteria X is harbored in the gut?

### B. Integrative analyses

Age-related diseases such as atherosclerosis cannot be explained by a single factor, like genetics alone, and in fact result from combinations of factors. Therefore, a large potential for discoveries and associated therapeutic opportunities is seen in integrative approaches that assess multiple factors together (Fig. 2 B) [1]. In addition, patients Electronic Health Records (EHR), as well as observations from model organisms (e.g. biomolecular "omics" data) generally have a sparse nature. For example, EHR data collection depends on patients condition and specific needs, thus different sets of data, measured in different sequences at different time points, are generally not available for each patient. Similarly, all studies have a different design, which prioritizes the investigation of certain tissues with different methods according to the scientific question at hand and available resources. In this context, observations across biological layers and studies may complement each-other: information that may be present, but could not be measured, in one biological layer, may be available from another mechanistically connected layer. For instance, genes (DNA), gene expression (RNA) and proteins are linked by the central dogma of cell biology: genes are transcribed into mRNA, which is translated into proteins. Proteins in turn form the backbone of metabolic pathways by acting on other biomolecules in complex cascades of biochemical reactions. Despite these links, observations from one layer are not sufficient to predict the state of other layers because of the many mechanisms of regulation that exist within and between these layers. However, integrating multiple related layers of biological data can offer a more informative picture than if considering each layer separately because they may complement each other while interactions between them could be also taken into account. Last but not least, integrative studies may highlight which type of observation, and which tissue, may carry the most relevant information with regard to a particular disease. This could help prioritizing experiments and guiding experimental designs to maximize the "return on investment" of generally expensive data collection processes.

Overall, and despite being collected under controlled genetic and environmental conditions, "omics" datasets derived from animal models of genetic diversity are complex, noisy and incomplete [14]. However, they still remain more comprehensive and coherent than their counterparts in human settings. While the need to overcome the limitations of pairwise associative approaches like GWAS is increasingly recognized, considering multiple layers of biological observations in integrative approaches remains challenging and to date only a few attempts were made in the field [1]. This is due to two main challenges: 1) the combination of heterogeneous and sparse data into coherent knowledge bases. And 2) deriving actionable insights out of complex patterns.

## V. OPEN SCIENCE AND DATA INTEGRATION

### A. Metadata are the glue that links datasets

Conceptually, the potential for the discovery of complex patterns grows with the heterogeneity and diversity of the analyzed data. But in order to keep results generalizable, the number of observations, or sample size, must grow as the diversity of considered observations increases (Fig. 2 C). Unfortunately, study budgets rarely allow to simultaneously collect large amount of diverse data from large cohorts. Instead, studies generally face a trade-off between the sample size and the number of observations and tissues collected, and prioritize these according to their scientific objectives. Combining data from multiple studies is therefore the solution to reach sample sizes that may not be achievable in a single study, and unlocking the investigation of new scientific questions. However, the construction of a knowledge base from multiple independent datasets greatly depends on the inter-operability of these datasets and essentially relies on the existence of sufficiently rich, detailed and standardized metadata (Fig. 2 D). While vertical integration (i.e. within a study) of datasets may be facilitated by the use of common nomenclatures and annotations within a same study, horizontal integration (i.e. across studies) is typically more challenging, as practices can differ between research groups.

### B. The FAIR principles

The FAIR principles were first formulated in 2016, as a mean to enable new discoveries by facilitating data integration and reuse [16]. These principles promote data management practices that enable the integration of compatible, or complementary, datasets. Indeed, Inter-operability and Reuse require that 1) data and metadata should be recorded using standard formats, notations and vocabularies, so that independent researchers could understand them and link information across datasets with as little ambiguity as possible. And 2), that the datasets should be documented with metadata that are rich and detailed enough for independent researchers to understand their exact provenance. In the case of integration, the description of the samples, the experimental design and the methods behind the data must allow any investigator to appreciate whether two datasets could be compared or merged together, and under which conditions.

### C. Data models

General metadata schema, such as schema.org (https://schema.org) or Dublin Core (https://dublincore.org) provide a generic tool to describe datasets in a standard manner, yet fall short in describing the complex context of experimental procedures behind most biomedical datasets. In biomedicine, domain-specific metadata schema such as the Investigation Study Assay (ISA) model provide an appropriate framework to link "omics" data through a database that describes their often intricate relationship of origin, measurement technology, sequencing runs, and experimental conditions [17]. Besides metadata schema, standard notations, controlled vocabularies and ontologies are essential to provide descriptions that can be searched and compared in an automated manner. Indeed, as the amount of generated data grows, so does the need to automate the process of metadata searching and matching.

### D. Driving forces

Public data repositories are instrumental in promoting good practices that facilitate data sharing and integration. For instance, domain-specific databases such as the European Nucleotide Archive (ENA) enforces the use of metadata models like ISA while generic repositories such as Zenodo promote more general-purpose standards like schema.org. Publishers and funding bodies increasingly demand that datasets associated to publications and projects are shared publicly on appropriate data sharing platforms. These strong driving forces in the research ecosystem facilitates the construction of knowledge bases across datasets and studies to enable larger integrative studies. However, data integration across independent "omics" studies remains challenging due to the inevitable differences and complexity of the experimental procedures.

### E. Reproducibility

Last but not least, documenting data processing is also critical to understand how processed data should be handled and interpreted as well as if and how they could be integrated. Although documenting data processing code has been facilitated by the now ubiquitous version control systems (e.g. Git), ensuring the actual reproducibility of data processing workflows remains a technical challenge to most biomedical researchers today. Indeed, reproducing workflows often demand, on top of the data processing code, specific sets of software dependencies (i.e. the computing environment) as well as an understanding of the links between processing steps, data sources and results (i.e. a knowledge graph of the workflow) [18]. This is typically addressed using virtualization technologies such as Docker (https://www.docker.com [19]) and workflow orchestration systems like the Common Workflow Language (CWL) [20]. Although building and using data processing systems that enable the full reproducibility of workflows can be perceived as an unessential overhead in today's context of competitive and time-pressured research, off-the-shelves complete data science technology stacks like Renku (https://renkulab.io [21]) are emerging. This will reduce the barriers to the adoption of technologies that enable the reproducibility of data processing, and facilitate data integration in the future.

## VI. EXPLAINABLE ARTIFICIAL INTELLIGENCE

### A. Machine learning unlocks integrative analysis

In biomedicine, machine learning (ML) is used or investigated in a variety of applications, from patient diagnosis and prognosis to the design of new drugs and the prediction of their effects [1]. It is a tool of choice to identify and use complex patterns across multi-modal data that are otherwise non-obvious to the human researcher and hard to assess with more classic statistical tools. While predictive algorithms have so far dominated this scene, there is a growing interest for methods including a strong explainability aspect. Indeed, in the context of aging research and systems genetics, which study the links between biomolecular factors and health-related traits, predicting phenotypes is typically of interest for disease-interception applications that require to anticipate the development of disorders in healthy individuals and applying preventive interventions to delay this development and expand healthspan. However, this is of little value for the identification of possible treatment targets, without an understanding of the key factors that drive the prediction. Interpretable ML methods that can highlight key predictive features of phenotypes that are relevant to diseases across multi-modal datasets are an emerging alternative to overcome the limitation of pairwise associative methods. An early example of such approach has been used to predict tissue-specific protein functions based on a network of protein-protein interactions built across a variety of tissues. The interpretable nature of the ML algorithm could then have been used to highlight the specific features important for the prediction of a function [22]. In the context of systems genetics, a similar method could be used to highlight, in a network of "omics" observations constructed across tissues, features relevant for the prediction of a trait of interest. A network could be based on gene co-expression, external knowledge bases such as publicly available annotations regarding gene-protein encoding (e.g. Ensembl [23]), protein-protein interactions (e.g. IntAct, MINT, MatrixDB [24]–[26]) and metabolic pathways (e.g. GO, KEGG [27]–[31]). Although building such networks across heterogeneous, noisy and sparse datasets requires significant efforts, such explainable Artificial Intelligence (XAI) framework could be a powerful tool to assist researchers in making discoveries.

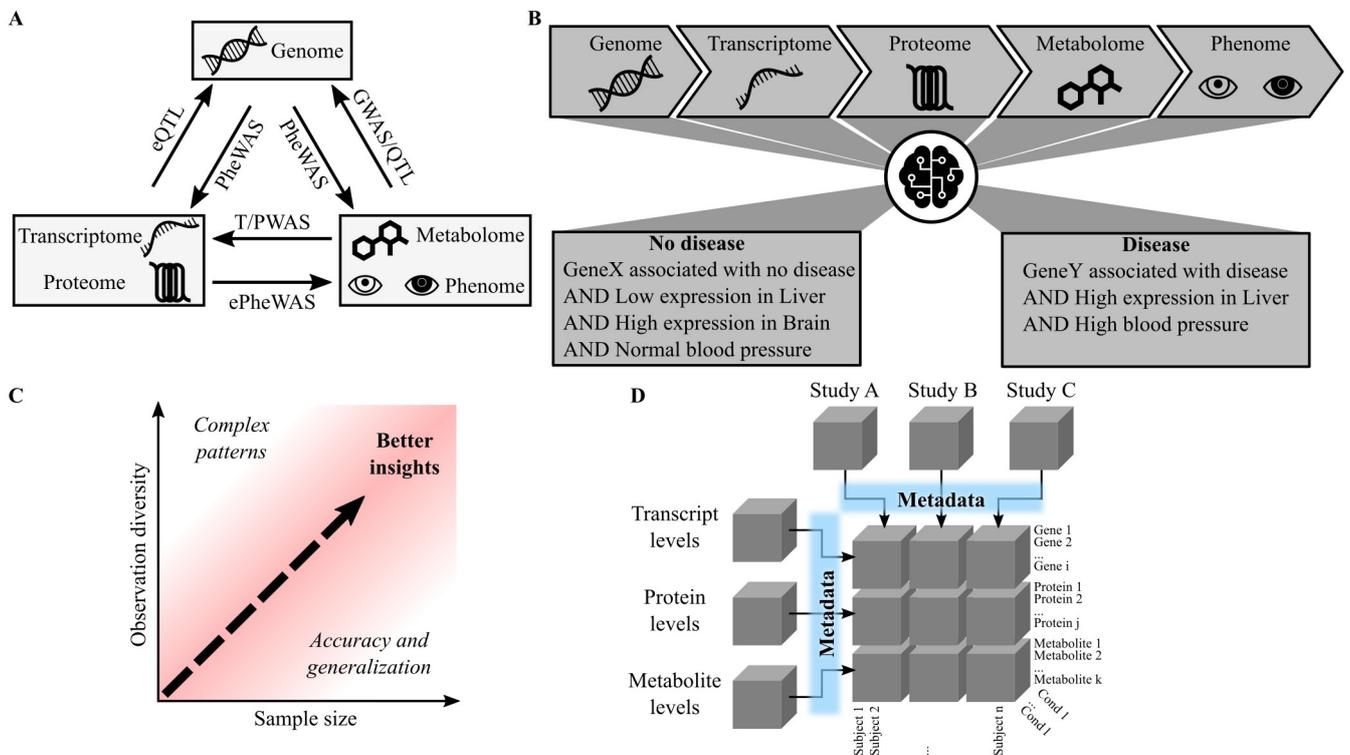

Fig. 2. **A)** Overview of "pairwise" associative studies in systems genetics. Genome Wide Association Studies (GWAS) and Quantitative Trait Locus (QTL) analyses search for associations between a phenotype and genetic variants. Phenome Wide Association Studies (PheWAS) look for associations with a gene variant within a collection of phenotypes or intermediate phenotypes. Expression Quantitative Trait Locus (eQTL) look for associations between an intermediate phenotype and genetic variants. Expression-based PheWAS (ePheWAS) search for associations with an intermediate phenotype within a collection of phenotypes. Transcriptome/Proteome-Wide Association Studies (T/PWAS) look for associations between a phenotype and variations in gene expression/protein levels. Adapted from Li *et al.* 2018 [14]. **B)** Integrative approaches assessing multiple layers of biological data together may capture complex patterns relevant to diseases, that could not be captured by "pairwise" associative approaches. Adapted from Li *et al.* 2018 [14] and Zitnik *et al.* 2019 [1]. **C)** The integration of diverse observations allows to capture more complex patterns, yet demand larger sample sizes in order to conserve the accuracy and generalization of insights. **D)** Vertical and horizontal data integration both require detailed and rich metadata. Metadata act as a glue that can link datasets across types of measurements (e.g. transcript, proteins or metabolite levels) or across studies.

VII. USE-CASE – MULTI-OMICS INTEGRATION IN SYSTEMS GENETICS

*A. The BXD family*

In a unique use-case, we intend to assemble a large knowledge base from a collection of "omics" and phenotype datasets collected on the BXD mouse genetic diversity panel and to use it to investigate integrative, exploratory approaches [32]. These datasets include genetics as well as gene expression, protein, lipids and metabolite levels measured across multiple tissues, as well as the composition of the gut microbiome and data from a large array of phenotyping tests targeting metabolic activity (blood pressure, body fat and lean mass, cardiac activity, glucose tolerance, etc.). This collection of datasets has been generated internally and its majority is publicly available on domain-specific repositories. Some of these data have been previously associated to discoveries reported across peer-reviewed publications [33]–[40]. These datasets are therefore well documented and our group has an excellent understanding of its complications and limitations, which is critical to determine under which conditions they can be integrated.

## B. Building a knowledge base

Our first goal is to consolidate a knowledge base through an extraction, transformation and load (ETL) process in order to standardize data and metadata notations across datasets and link measurements across strains, tissues, experimental conditions and assays technologies (i.e. vertical integration). The use of domain-specific ontologies and standards, such as the Mammalian Phenotype ontology (MP) (http://purl.obolibrary.org/obo/mp.owl), the Vertebrate Trait ontology (VT) (http://purl.obolibrary.org/obo/vt), the Mouse Adult Gross Anatomy (MA) (http://purl.obolibrary.org/obo/ma.owl) and the Ontology of Biomedical Investigations (OBI) (http://purl.obolibrary.org/obo/obi [41]) will also facilitate future integration with independent studies carried out internally or by other research groups (i.e. horizontal integration).

## C. Machine learning on graphs

This knowledge base will be used as a test bed for the development of integrative data analysis approaches based on ML. This will primarily focus on approaches based on graphs, as these are tools of choice to describe heterogeneous biological observations together with their links (such as interactions between proteins, mechanistic links between genes, transcripts and proteins and metabolic pathways). In particular, we envision novel applications for graph Convolutional Neural Networks (CNN) [42]–[45]: Traits relevant to diseases could be predicted based on a network of connected "omics" observations across tissues. The inherent interpretability of graph convolutional neural networks could then highlight key predictive features of this network, which would help discovering complex biological mechanisms and potential therapeutic targets (Fig. 3).

## VIII. CONCLUSION

There is a critical need to develop new strategies for preventing, delaying or reversing the course of age-related diseases in the growing and aging global population. Age-related diseases have complex multi-factorial roots which demand to take individual physiological characteristics into consideration both for research and treatment. Understanding the metabolic mechanisms underlying the aging process helps developing interventions to compensate its effects. In particular, integrative approaches that combine biological observations across multiple tissues promise to generate valuable insights into these complex biomolecular mechanisms. Although "omics" technologies and the use of model organisms enable the detailed investigation of tissues and cells physiology, identifying complex patterns and regulatory systems across tissues and biomolecular layers remains challenging. Indeed, it requires integrative approaches that can combine a large amount - and a diversity - of "omics" observations across multiple tissues and varying conditions. Integrative analyses need to combine multiple datasets of different types (i.e. genomic, proteomic, metabolomic) that could be generated within a same, or within multiple independent studies. Such integration requires a deep understanding of each dataset's characteristics and specificities. This demands rich, detailed and harmonized documentation and metadata. Although promoted by increasingly adopted open science standards, the necessary level of details is rarely accessible for publicly available datasets and such approach therefore remain marginal in biomedicine.

In order to provide a first use-case in the field of systems genetics, we are assembling a large knowledge base of heterogeneous "omics" datasets derived from the BXD mouse genetic diversity panel. This will enable to test the application of XAI methods for assisting researchers in the discovery of complex biological mechanisms relevant for age-related diseases. This study will allow to investigate the links between genetics, metabolism, tissues and phenotypes. It may enable the identification of novel therapeutic targets against complex disorders and set the ground for further integrative approaches in biomedicine.

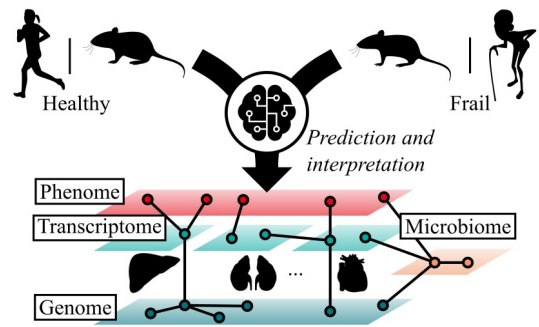

Fig. 3. Integrative systems genetics approach based on graphs. Biological data layers including diverse types of observations (e.g. genetic variations, gene transcription, phenotypes or microbiome composition) across multiple tissues (e.g. liver, kidney, heart) are integrated into a graph. Graph features that are key to predict a trait that is relevant for a disease (e.g. the body mass) are extracted using the interpretability of a graph-based ML algorithm.

## ACKNOWLEDGMENT

We thank Benjamin Ricaud (EPFL) for his advises and guidance regarding graph-based ML, Christine Choirat (Swiss Data Science Center - SDSC), Eric Bouillet (SDSC) and Emma Jablonski (SDSC) for their advise and support regarding reproducible data science. We thank the SDSC for funding this work as part of its collaborative data science program (project SysGen).

## GLOSSARY

- **BXD mouse genetic diversity panel**: A set of ~200 strains of recombinant inbred mice derived from C57BL/6 and DBA/2 parents. Thanks to patterns of genetic recombinations that are unique to each inbred strain, this family of mice allows to resolve the effect of 6 million DNA variants on heritable traits.

- **FAIR principles**: Data management guidelines formulated in the 2016 paper "The FAIR Guiding Principles for scientific data management and stewardship" by Wilkinson et al. aiming at promoting Findability, Accessibility, Interoperability, and Reuse of digital objects. The FAIR principles constitute a key stone in open science as they clearly identify the essential elements needed for data reuse by the community.

- **Healthspan**: The period of life in which a person is in healthy condition.

- **Explainable Artificial Intelligence (XAI) and interpretable machine learning (ML)**: ML approaches focusing on models which underlying logic can be understood by the user. Explainable ML models are seen as white boxes. In contrast, models which logic cannot be understood by the user because it is based on too high levels of abstraction are considered black boxes.

- **Knowledge base**: In this text, the term "knowledge base" is used in its most general sense to describe any form of organized information around a dataset, indistinctively of its form or complexity (i.e. whether as a simple text table or as a complex relational or graph database). This includes metadata, metadata description and possible links within and between these elements.

- **Precision medicine**: An approach to medicine that takes patient individual characteristics into account for the design of personalized treatments. While this concept is not new and has been applied in the past (e.g. blood transfusion needs to be adapted to patient's blood type), the terms "precision medicine" (interchangeable with the term "personalized medicine") refer to emerging approaches that account for complex characteristics, or combinations of them, found in genetics, life-style and a patient's environment.

- **Systems genetics**: A research approach to understand complex traits. Systems genetics investigates the links between genetics variations, intermediate molecular phenotypes (i.e. gene expression, metabolites levels, etc.) and traits.